\documentclass[runningheads]{llncs}
\usepackage{graphicx}

\usepackage{hyperref}

\usepackage{amsmath}
\usepackage[linesnumbered,ruled]{algorithm2e}
\usepackage{verbatim}
\usepackage{xcolor}
\usepackage{tabularx}
\usepackage{float}
\usepackage{array}
\usepackage{multirow}
\usepackage{multicol}
\usepackage{lineno}
\usepackage{verbatim}
\usepackage{rotating}
\usepackage{subcaption}
\usepackage[title]{appendix}

\begin{document}

\title{\textit{Shackled}: a 3D Rendering Engine Programmed Entirely in Ethereum Smart Contracts}

%
\titlerunning{Shackled: On-chain 3D Rendering with Ethereum}

%
\author{
Ike Smith\inst{1} \and
Casey Clifton\inst{1,2}
}
\authorrunning{Ike Smith \& Casey Clifton}
%
\institute{
$^1$Spectra, Australia \\
\href{www.spectra.art}{www.spectra.art}\\
\email{ike@spectra.art}, \email{caseyclifton@proton.me}
}

\maketitle              
\footnotetext[2]{Corresponding author.}
%
%



\begin{abstract}


The Ethereum blockchain permits the development and deployment of smart contracts which can store and execute code ``on-chain'' --- that is, entirely on nodes in the blockchain's network. Smart contracts have traditionally been used for financial purposes, but since smart contracts are Turing-complete, their algorithmic scope is broader than any single domain. To that end, we design, develop, and deploy a comprehensive 3D rendering engine programmed entirely in Ethereum smart contracts, called Shackled. Shackled computes a 2D image from a 3D scene, executing every single computation on-chain, on Ethereum. To our knowledge, Shackled is the first and only fully on-chain 3D rendering engine for Ethereum. In this work, we 1) provide three unique datasets for the purpose of using and benchmarking Shackled, 2) execute said benchmarks and provide results, 3) demonstrate a potential use case of Shackled in the domain of tokenised generative art, 4) provide a no-code user interface to Shackled, 5) enumerate the challenges associated with programming complex algorithms in Solidity smart contracts, and 6) outline potential directions for improving the Shackled platform. It is our hope that this work increases the Ethereum blockchain's native graphics processing capabilities, and that it enables increased use of smart contracts for more complex algorithms, thus increasing the overall richness of the Ethereum ecosystem.




\end{abstract}

\section{Introduction}

The Ethereum blockchain has often been described as a ``World Computer'' \cite{buterin2013ethereum} as it offers a smart contract platform that allows for general algorithms to be ran ``on-chain'' --- that is, entirely on nodes in the blockchain's network. Smart contract code has interesting and novel benefits \cite{zheng2020overview}, namely; immutability (records on Ethereum cannot be edited once committed to a block), permanance (the code lasts as long as the blockchain remains in existence), transparency (all smart contract source code is visible publically), scalability (smart contract code can be executed on any node in the network), exact reproducibility (other programs can execute the code identically given the same inputs), and composability (other programs can use the code as a building block in a larger system). 

In this work, we program a 3D rendering engine --- that is, a program which takes as input a three dimensional scene, and computes as an output a two dimensional image representing what that scene would look like from a given camera pose --- entirely within Solidity smart contracts. We then deploy it on the Ethereum blockchain. We name the engine \textit{Shackled} due to its on-\textit{chain} nature. To the best of our knowledge, \textbf{Shackled is the first fully on-chain 3D rendering engine}.

The contributions of this work are as follows:
\begin{enumerate}
    \item Quantification of the efficiency of Shackled in terms of gas usage. 
    \item \textit{Three} 3D object datasets that are compatible with Shackled.
    \item A demonstration of the utility of Shackled: we develop and deploy a generative art project (\textit{Shackled Genesis}), integrating Shackled with non-fungible tokens (NFTs). 
    \item A user interface to the {Shackled} smart contract library called \textit{Shackled Creator}\footnote{Available at \href{https://shackled.spectra.art/\#/creator}{https://shackled.spectra.art/\#/creator.}}, which allows users to render 3D objects entirely on the blockchain through a simple, no-code, visual interface on the web.
    \item A complete discussion regarding the challenges of programming complex algorithms as Solidity smart contracts.
\end{enumerate}





\section{Related work}

\textbf{Solidity smart contracts} are a relatively young technology \cite{dannen2017introducing}, and have been used for primarily financial use cases. Perhaps most notably, enabling the development of Decentralised Finance (DeFi); a class of finance characterised by blockchain-based trustlessness and secure peer-to-peer transactions in the absence of a trusted third party \cite{collibus2021role}. However, the applications of smart contracts extend well beyond the domain of finance, with on-chain implementations of complex algorithms appearing in the Ethereum ecosystem (e.g., an on-chain chess engine \cite{fiveoutofnine2021onchainchess}).

One novel yet controversial application of smart contracts is the \textbf{tokenisation of art} NFTs \cite{wang2021non,chohan2021non}. NFTs have been criticised for their implications on traditional copyright law \cite{goldman2021non}, breaches of intellectual property \cite{yoder2022opensea}, effect on art \cite{frye2021nftsdeathofart}, environmental impact \cite{truby2022blockchain} (this is discussed in the context in this project in Appendix~\ref{s:carbon}), over-hyped status in the modern in zeitgeist \cite{mackenzie2021nfts}, and their lack of regulation promoting a predatory culture of scamming, manipulation, and fraud \cite{twomey2020fraud}. 

Conversely, they have also been praised for redefining the nature of ownership in the digital age \cite{wang2021non,chohan2021non}, status as a novel investment opportunity, and for the \textit{positive} effect that they have had on art culture \cite{kugler2021future} and collectorship. 

Much work has been done regarding the tokenisation of art, including Ethereum-based generative art projects \cite{cxkoda2021sa,divergence2021brotchain,nouns2021}, on-chain graphics libraries \cite{wattsy2021kohi}, storing of 3D objects on-chain \cite{sayangel2022blitbox}, and using on-chain technology for the generation and preservation of digital art \cite{autoglyphs2019,cryptopunks2017,proof2022moonbirds}.

At the intersection of art tokenisation and smart contract programming is the potential for performing \textbf{3D rendering} on-chain. Approaches to decentralised 3D rendering are being explored and commercialised \cite{rendertoken2017}, with a focus on large-scale computing on custom blockchains by leveraging graphs of computing nodes \cite{alchemy2022,ward2020practical} optimised for graphics processing. These approaches transfer the hardware burden from the user to a decentralised network of compute providers, essentially creating a market for specialised computation.

Although Solidity smart contracts represent a great opportunity for algorithmic development, the scale of computation that they allow at this time is limited; modern graphics engines could not currently be implemented in Solidity. As such, Shackled is based on the work of early graphics pioneers, using technology from nearly $50$ years ago (which is more suited for implementation on-chain today). Specifically, we modify versions of Bui Tong Phuong's \cite{phong1975illumination} and Jim Blinn's \cite{blinn1977models} original 3D rendering and lighting models, and use them to create a Solidity version of a simple rendering pipeline inspired by OpenGL \cite{segal1999opengl}.





\section{Datasets}


We begin by developing and presenting three datasets in Table~\ref{fig:datasets} for use with Shackled which will be used throughout this work (see Appendix~\ref{s:dataavail} for data availability). Firstly, the \textbf{Shackled Genesis Dataset}, which comprises $1024$ triangular prism-based geometries generated using an on-chain geometry generating algorithm. Secondly, the \textbf{Shackled Icons Dataset}, which is comprised of a series of hand-crafted 3D objects. Thirdly, the \textbf{Shackled Common Graphics Objects Dataset}, which is comprised of common objects from 3D graphics history, processed for compatibility with Shackled.

\section{Approach}


\subsection{Development}

\label{s:dev}

Shackled was developed using common development tools for programming and testing Solidity smart contracts on Ethereum. Our \textbf{technical stack} included \textit{Hardhat} as the primary development environment, \textit{Ethers} for interfacing with the smart contracts, and \textit{Chai} for testing. We catalogue the development challenges and how they were overcome in Table~\ref{tab:issues}.

\begin{figure}[H]
    \centering
    \setlength{\tabcolsep}{0pt}
    \begin{subfigure}{1.0\textwidth}
        \begin{tabular}{cccccc}
            \includegraphics[width=0.16666666666\linewidth]{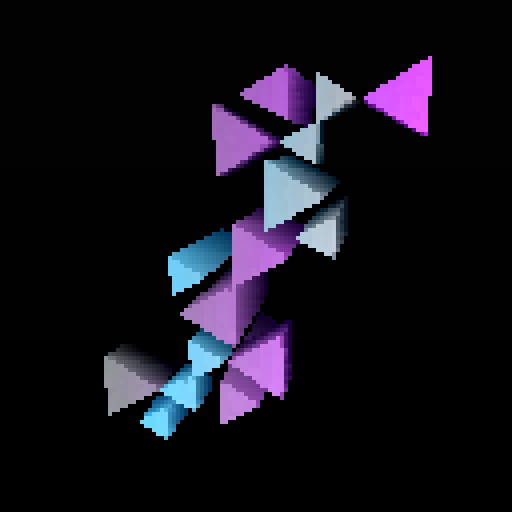}
            & 
            \includegraphics[width=0.16666666666\linewidth]{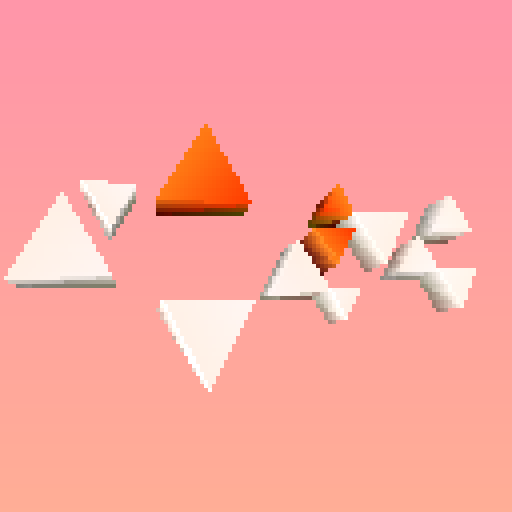}
            & 
            \includegraphics[width=0.16666666666\linewidth]{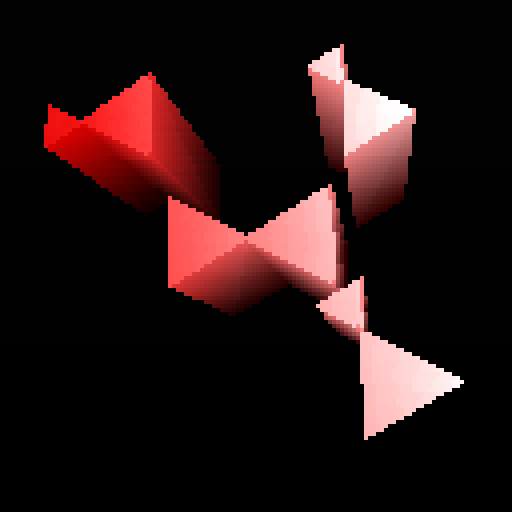}
            & 
            \includegraphics[width=0.16666666666\linewidth]{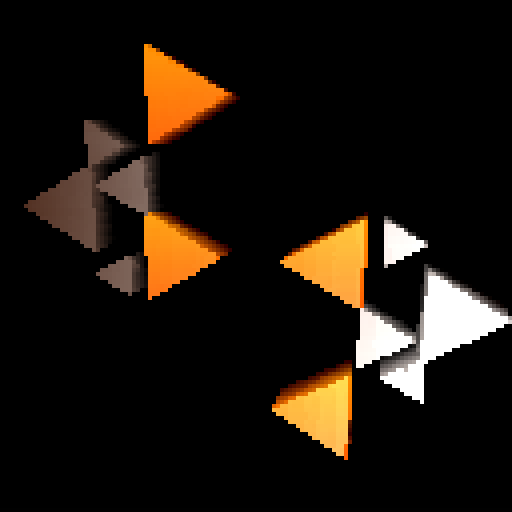}
            & 
            \includegraphics[width=0.16666666666\linewidth]{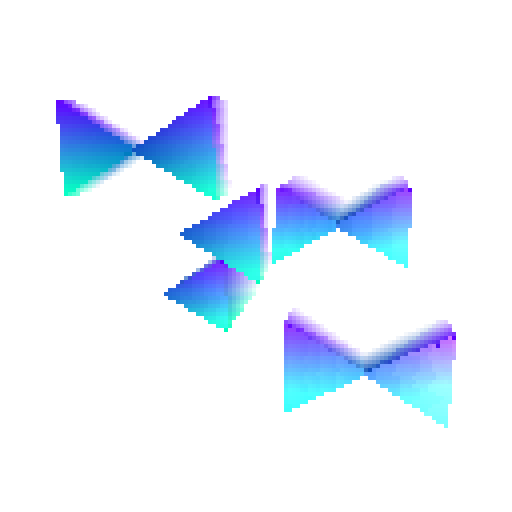}
            & 
            \includegraphics[width=0.16666666666\linewidth]{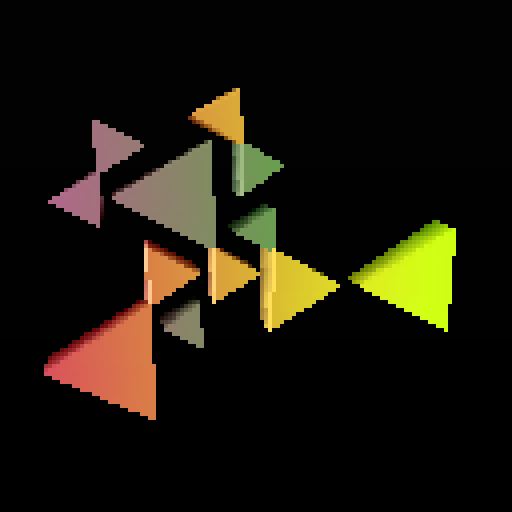}\\[-4pt]
            
            \includegraphics[width=0.16666666666\linewidth]{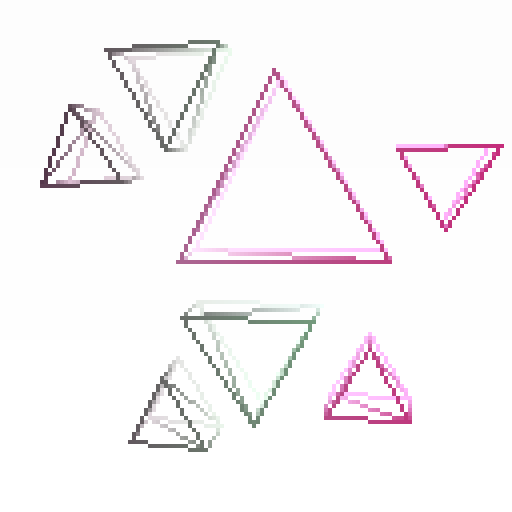}
            & 
            \includegraphics[width=0.16666666666\linewidth]{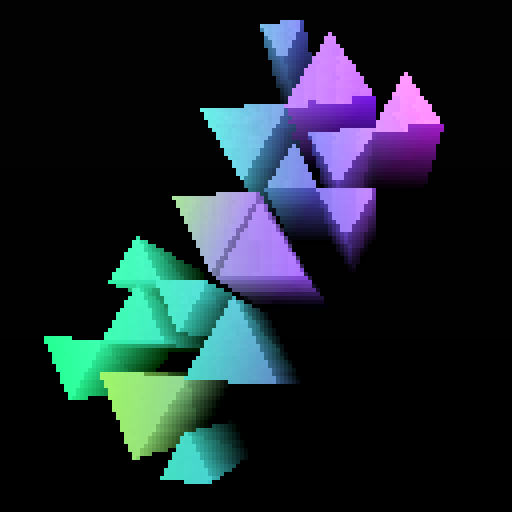}
            & 
            \includegraphics[width=0.16666666666\linewidth]{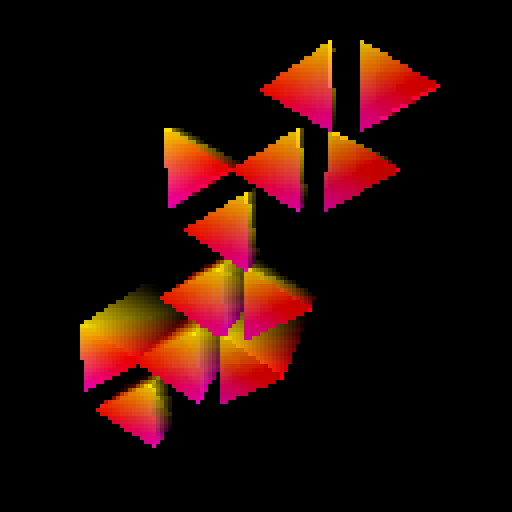}
            & 
            \includegraphics[width=0.16666666666\linewidth]{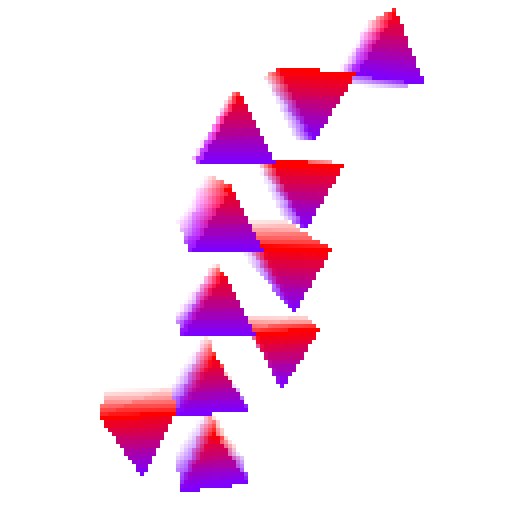}
            & 
            \includegraphics[width=0.16666666666\linewidth]{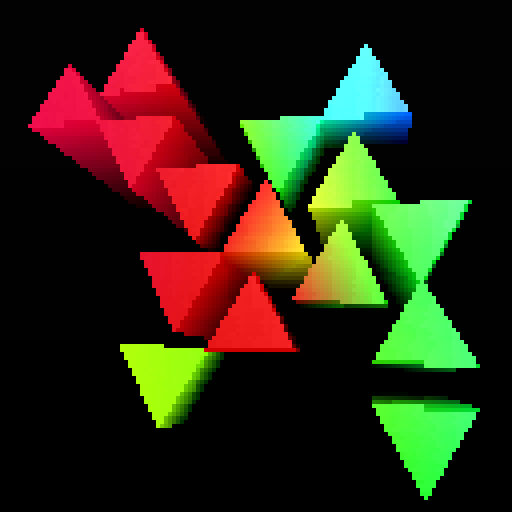}
            & 
            \includegraphics[width=0.16666666666\linewidth]{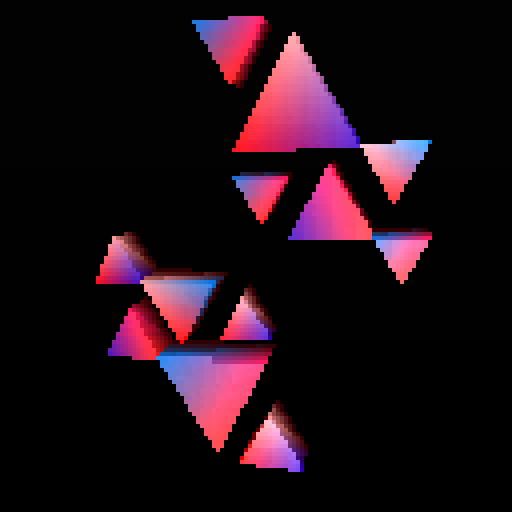}\\[-4pt]
            
        \end{tabular}%
        \caption{Examples from the \textbf{Shackled Genesis Dataset}, as rendered in Shackled. Each instance features a number of triangular prisms placed generatively and rendered with different colour and lighting parameters. There are $1024$ instances in the dataset.}
    \end{subfigure}
    \begin{subfigure}{1.0\textwidth}
        \begin{tabular}{cccccc}
            \includegraphics[width=0.16666666666\linewidth]{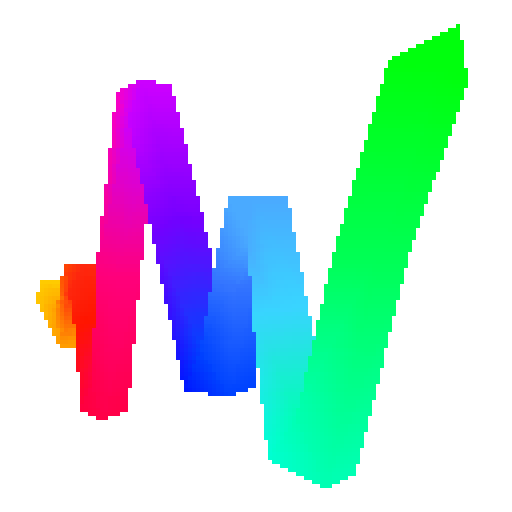}
            & 
            \includegraphics[width=0.16666666666\linewidth]{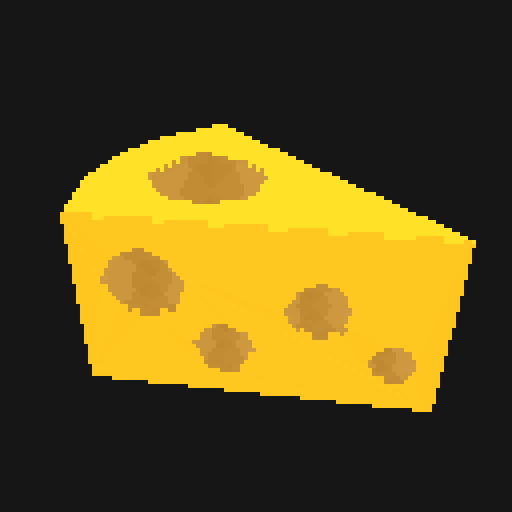}
            & 
            \includegraphics[width=0.16666666666\linewidth]{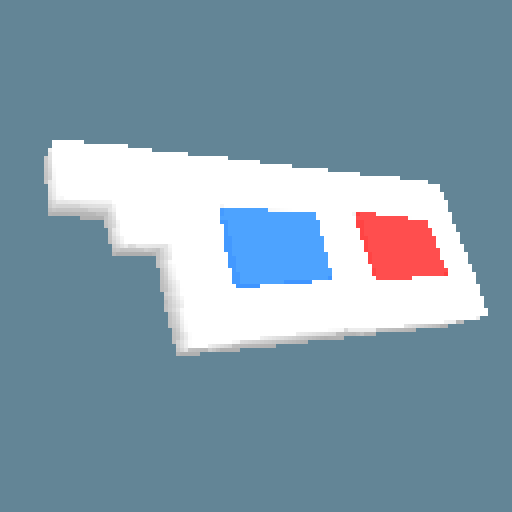}
            & 
            \includegraphics[width=0.16666666666\linewidth]{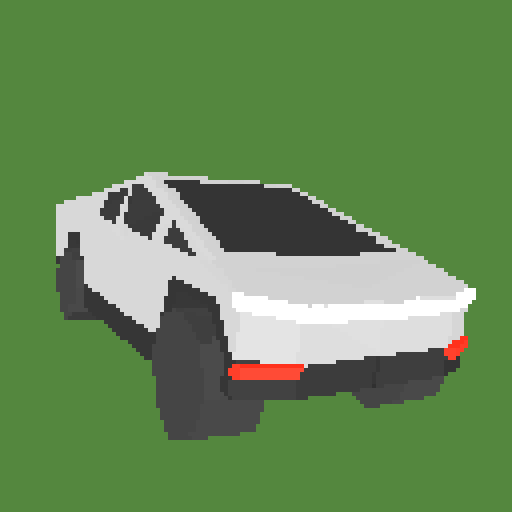}
            & 
            \includegraphics[width=0.16666666666\linewidth]{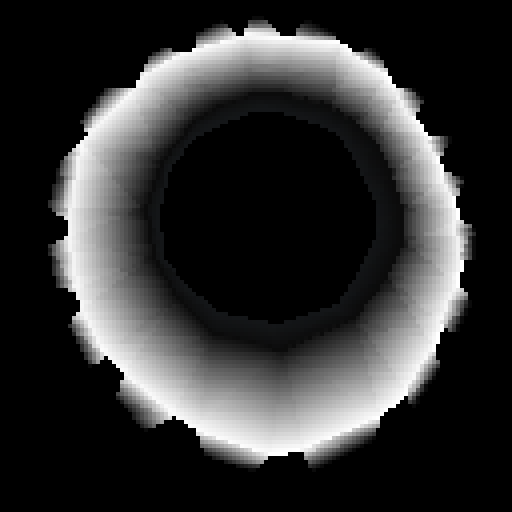}
            & 
            \includegraphics[width=0.16666666666\linewidth]{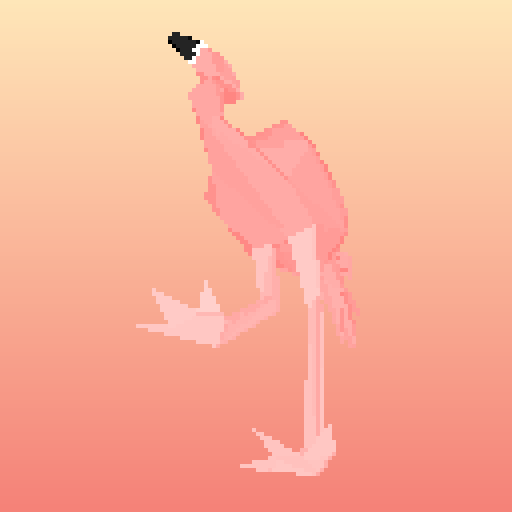}\\[-4pt]
            
            \includegraphics[width=0.16666666666\linewidth]{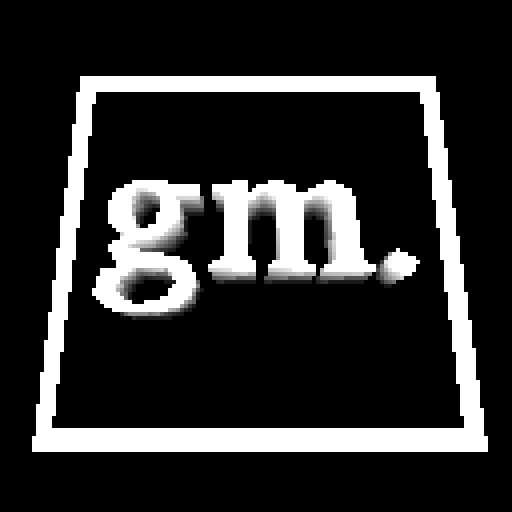}
            & 
            \includegraphics[width=0.16666666666\linewidth]{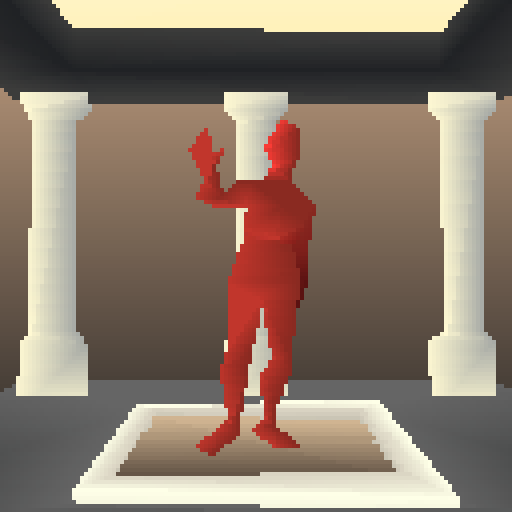}
            & 
            \includegraphics[width=0.16666666666\linewidth]{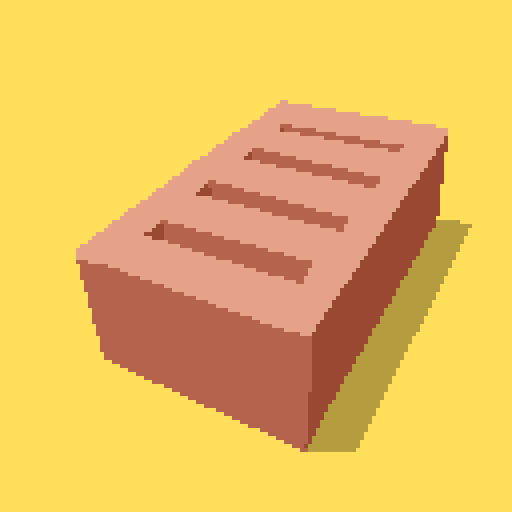}
            & 
            \includegraphics[width=0.16666666666\linewidth]{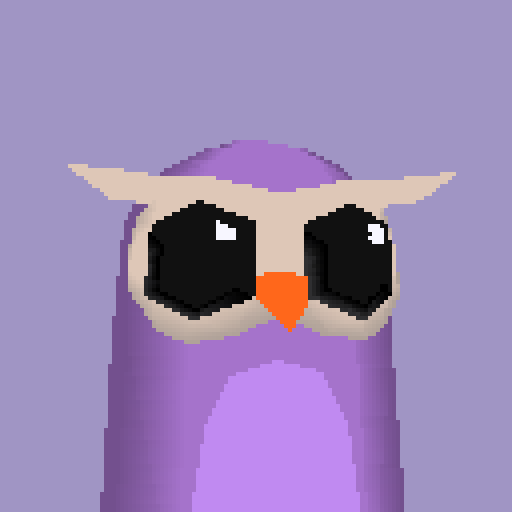}
            & 
            \includegraphics[width=0.16666666666\linewidth]{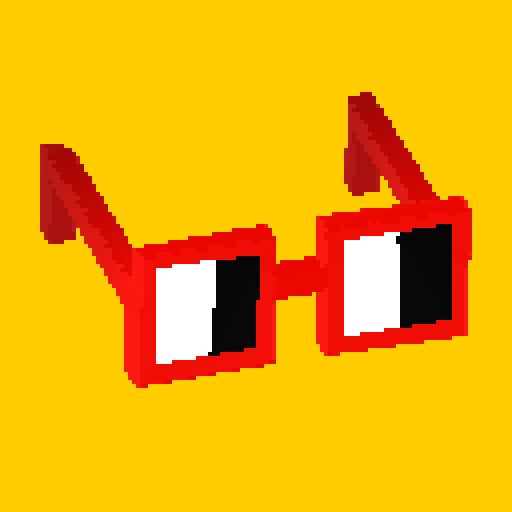}
            & 
            \includegraphics[width=0.16666666666\linewidth]{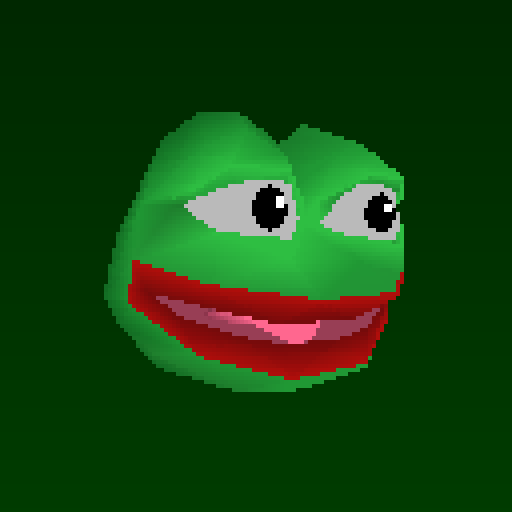}\\[-4pt]
        \end{tabular}%
        \caption{Examples from the \textbf{Shackled Icons Dataset}, as rendered in Shackled. The subjects of each render were chosen for recognisability and relevance in the on-chain art and NFT community.}
    \end{subfigure}
    \begin{subfigure}{1.0\textwidth}
        \begin{tabular}{cccccc}
            \includegraphics[width=0.16666666666\linewidth]{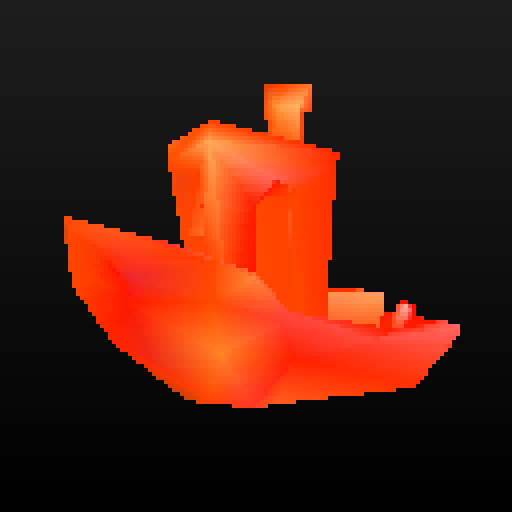}
            & 
            \includegraphics[width=0.16666666666\linewidth]{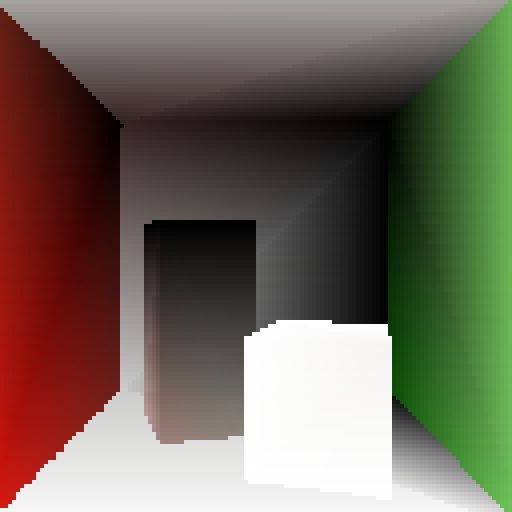}
            & 
            \includegraphics[width=0.16666666666\linewidth]{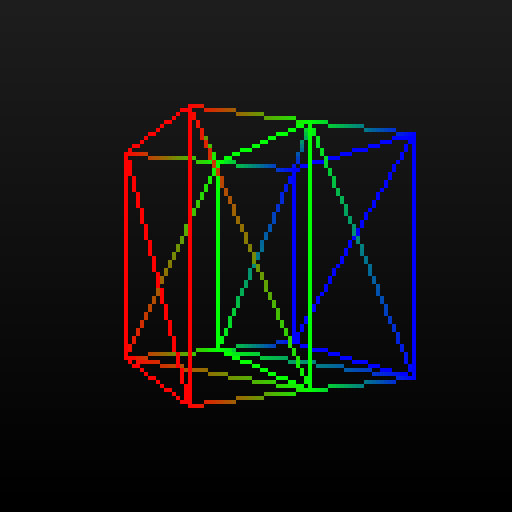}
            & 
            \includegraphics[width=0.16666666666\linewidth]{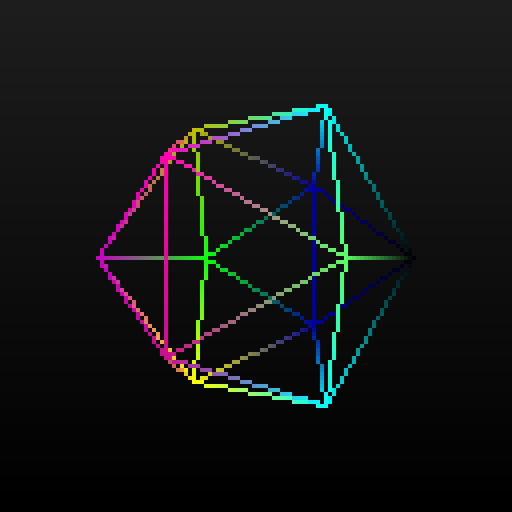}
            & 
            \includegraphics[width=0.16666666666\linewidth]{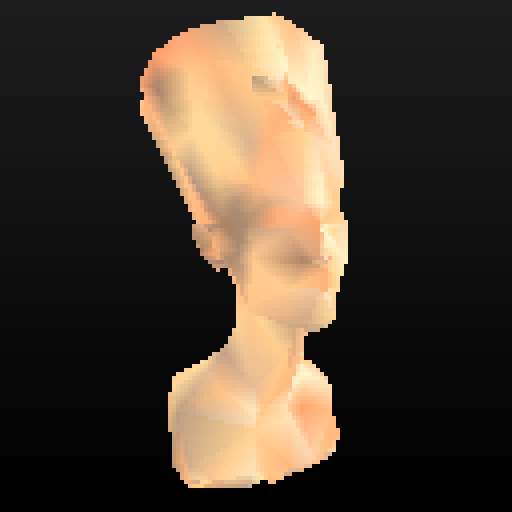}
            & 
            \includegraphics[width=0.16666666666\linewidth]{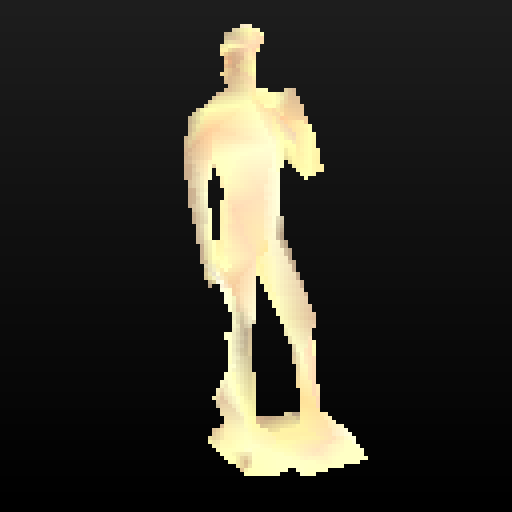}\\[-4pt]
            
            \includegraphics[width=0.16666666666\linewidth]{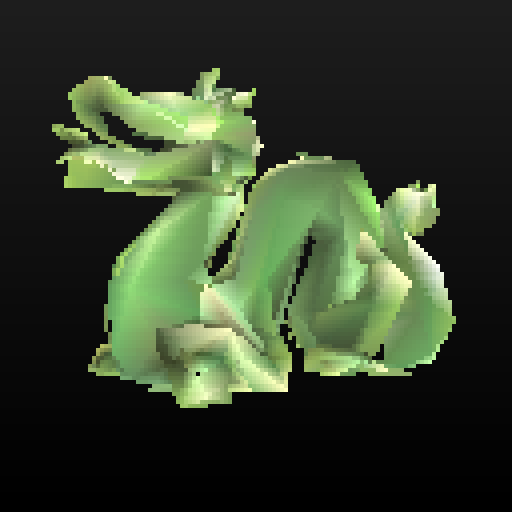}
            & 
            \includegraphics[width=0.16666666666\linewidth]{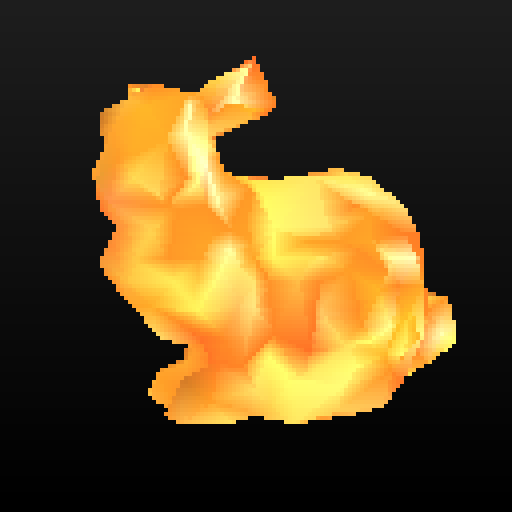}
            & 
            \includegraphics[width=0.16666666666\linewidth]{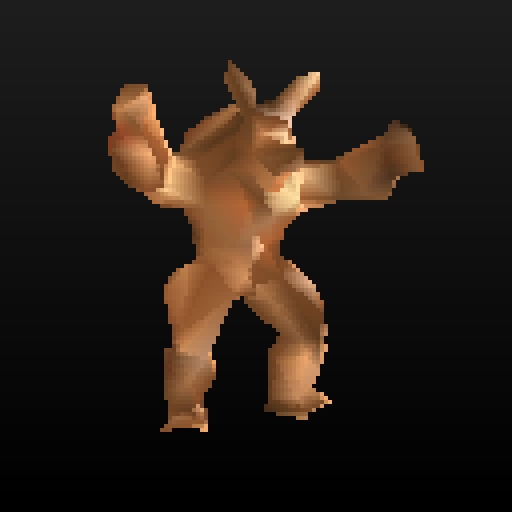}
            & 
            \includegraphics[width=0.16666666666\linewidth]{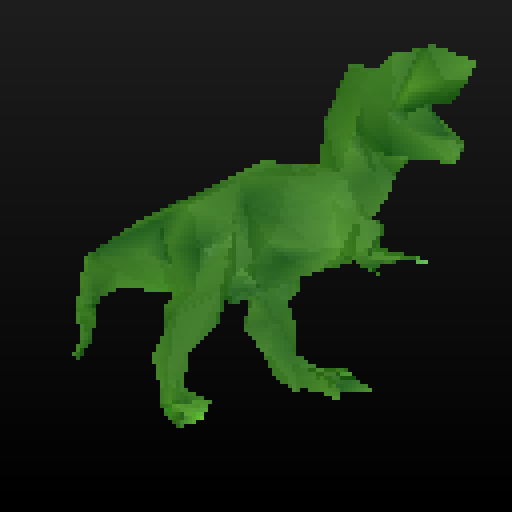}
            & 
            \includegraphics[width=0.16666666666\linewidth]{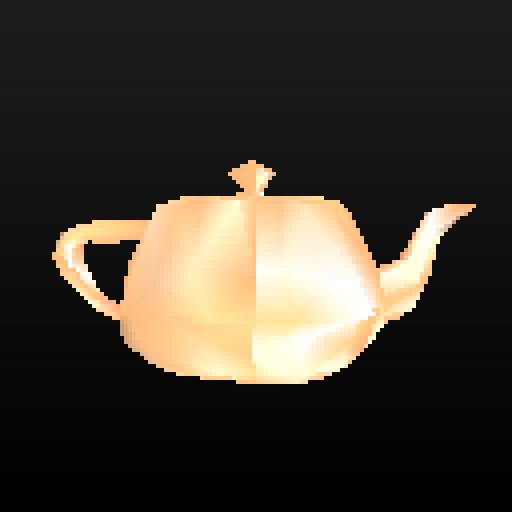}
            & 
            \includegraphics[width=0.16666666666\linewidth]{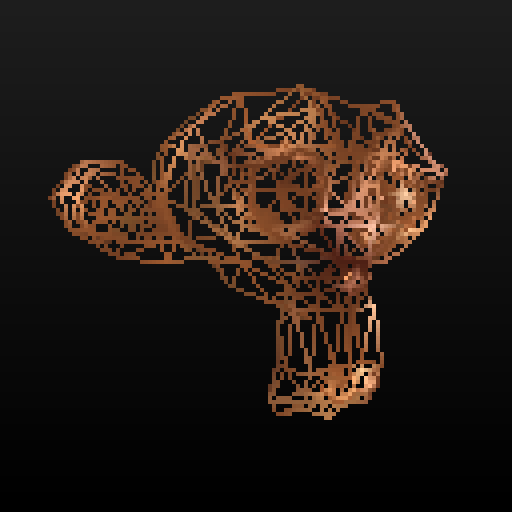}\\[-4pt]
        \end{tabular}%
        \caption{Examples from the \textbf{Shackled Common Graphics Object Dataset}, as rendered in Shackled. \textbf{Top row}: 3DBenchy (3D printing benchmark), Cornell's box \cite{goral1984modeling}, wireframe renders of a Cube and an Isosphere, a 3D scan of the bust of Nefertiti \cite{wiedemann1982bust}, and a 3D scan of Michaelangelo's sculpture David \cite{levoy2000digital}. \textbf{Bottom row}: Stanford's Dragon, Bunny, Armadillo, and Tyra \cite{stanford2014scanning}, the Utah Teapot, and Blender's Suzanne \cite{blain2019complete}.}
    \end{subfigure}
    \caption{Datasets used in this work.}
    \label{fig:datasets}
\end{figure}

\begin{table}[H]
    \centering
    \caption{Descriptions of development challenges and the solutions used to overcome them during the development of Shackled. We categorise the challenges as being imposed by Solidity, or by the Ethereum ecosystem.}
    \begin{tabular}{|p{6cm}|p{6cm}|}
        \hline
        \multicolumn{1}{|c|}{Description} & \multicolumn{1}{|c|}{Solution} \\
        \hline
        \multicolumn{2}{|c|}{\textit{Solidity programming language challenges}} \\
        \hline
        No support for floating point or fixed-point numbers. &  Use whole numbers in the gwei designation ($10^9$). \\
        \hline
        Data types less than 32 bytes being assembled into the same storage slot, reducing division precision.  &  Pad variables to ensure they take up $32$ bytes, and use unit tests to ensure correct behavior. \\
        \hline
        Overflow and underflow errors are common as integers are a fixed size. &  The latest versions of Solidity include automatic checks for these errors. \\
        \hline
        Limited best-practice documentation. & Learn from exemplar projects (all on-chain code is open source). \\
        \hline
        Uninformative stack trace errors. & use development environments such as Hardhat.  \\
        \hline
        Limited / no typecasting. &  Develop workarounds for types that can't be cast to other types. There is not necessarily a `catch-all' solution to this issue. \\
        \hline
        Only 16 variables being usable at a time due to the stack depth limit. &  Use structs to store and pass variables. \\
        \hline
        \multicolumn{2}{|c|}{\textit{Ethereum ecosystem challenges}} \\
        \hline
        Exceeding gas limits imposed by node providers. & Run your own node or find a provider willing to provide a higher limit (Alchemy in our case). \\
        \hline
        Modifying / writing storage incurs significant expense on the Ethereum blockchain. & No general solution, but on-chain compression and generation can be useful to avoid directly storing data. \\
        \hline
        True randomness is not possible as Ethereum is deterministic. &  Psuedorandomness can be achieved by using a block's hash or number as input. \\
        \hline
        The EVM stack has only 1024 slots available for functions calling other functions. & Use loops in lieu of recursion. \\
        \hline
    \end{tabular}
    \label{tab:issues}
\end{table}

To benchmark Shackled and \textbf{estimate the cost of computation}, we estimate the amount of \textbf{gas}; a unit describing the amount of computational power required to perform some specific computation on the Ethereum network \cite{buterin2013ethereum}. Our {gas estimation approach} involves binary searching the upper and lower gas limits until a `not enough gas' error is triggered; we repeatedly attempt to complete renders with less and less gas until we converge on the true gas required to perform a specific render operation, allowing us to accurately measure the computational cost of completing a given render. 

\textbf{Importantly, Shackled does not require the expenditure of gas to perform rendering operations.} The entire rendering operation is implemented in a \textit{read call}, and thus does not \textit{write} any data to the Ethereum blockchain. \textbf{As such, all gas estimations presented in this study are \textit{gas-equivalents}}, e.g., a Shackled render benchmarked at $10$ billion units of gas does not actually consume that amount of gas, but its computational cost is equivalent to an operation that would consume that amount. 


\subsection{Design of the rendering pipeline}

\begin{figure}[H]    
    \centering
    \setlength{\tabcolsep}{0pt}
    \begin{tabular}{cccccc}
        \includegraphics[width=0.16666666666\linewidth]{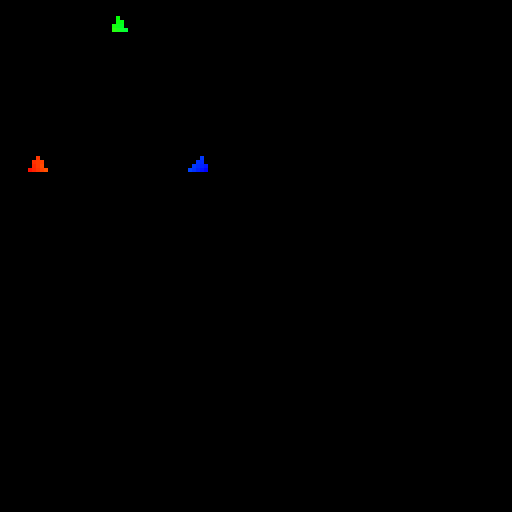}
        & 
        \includegraphics[width=0.16666666666\linewidth]{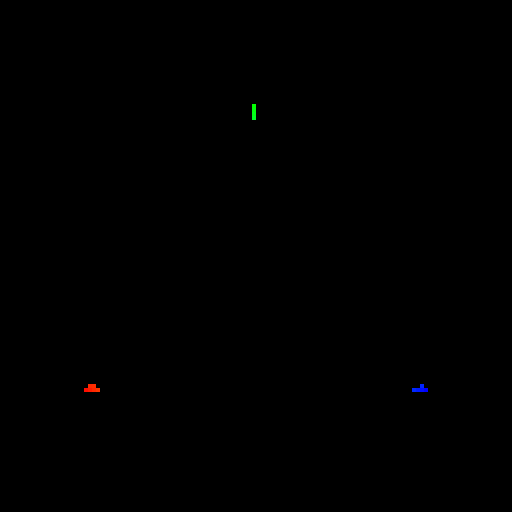}
        & 
        \includegraphics[width=0.16666666666\linewidth]{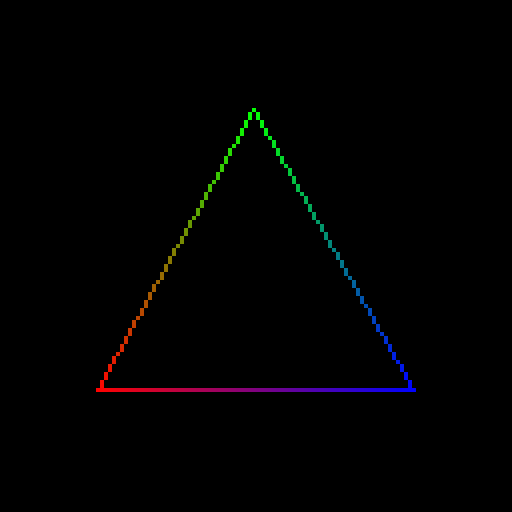}
        & 
        \includegraphics[width=0.16666666666\linewidth]{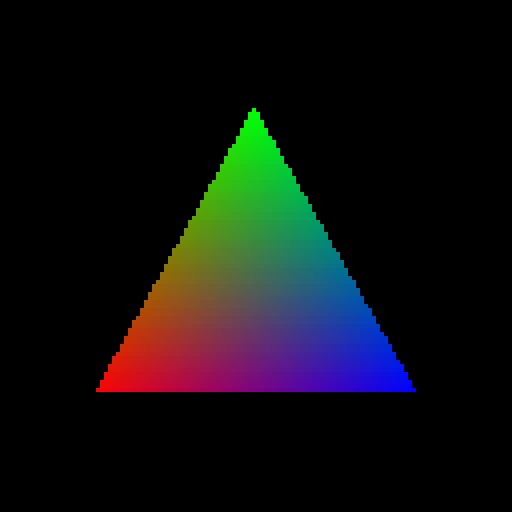}
        & 
        \includegraphics[width=0.16666666666\linewidth]{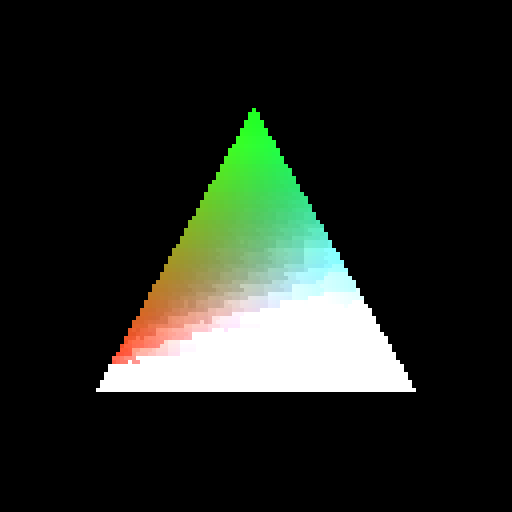}
        & 
        \includegraphics[width=0.16666666666\linewidth]{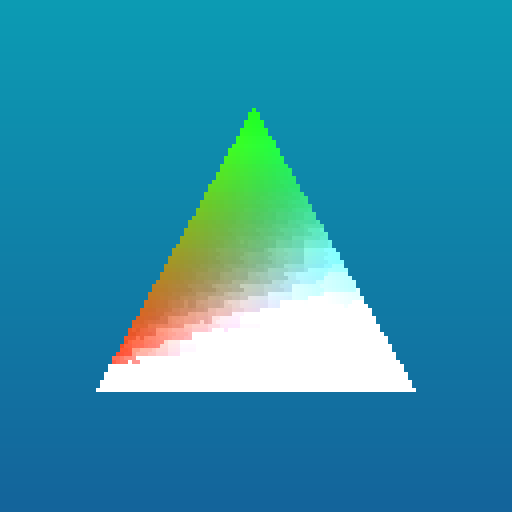}\\[-4pt]
    \end{tabular}%
    \caption{The results of the key steps of the Shackled rendering pipeline for rending a single triangle. \textbf{From left to right}, we display the outputs of 1) vertex specification, 2) the vertex shader, 3) primitive assembly, 4) rasterization, 5) the fragment shader, 6) compositing and image buffer return. Ultimately the render comprises of a red-green-blue coloured triangle light with a harsh specular light with a steep fall-off from the bottom right, composited onto a blue gradient background. All steps are computed on the Ethereum blockchain entirely within smart contracts.}
    \label{fig:pipeline}
\end{figure}

Shackled converts a 3D model into a 2D image using a sequential rendering pipeline inspired by OpenGL \cite{segal1999opengl} and modified for on-chain implementation and execution (see Figure~\ref{fig:pipeline}). The key steps are as follows.

\textbf{Vertex specification} involves providing the 3D positions and colours for all points in the object's mesh. Faces will be constructed from these 3-vectors as per the {\fontfamily{qcr}\textit{.obj}} file format specification \cite{mchenry2008overview}.

The \textbf{vertex shader} is traditionally a program written in a language designed for accelerated computation on graphics hardware, which allows for parallel processing of independent vertex computations. In Shackled, we have no such accelerative capability. The vertex shader still assumes the traditional role of projecting the points using a perspective or orthographic camera matrix, and transforming the points via a model view matrix (translations and scaling are supported), converting the points to world space.

\textbf{Primitive assembly} involves taking groups of 3 vertices and constructing triangles out of them (Shackled only supports triangulated objects as input). The vertex indices of the triangles are specified during vertex specification. No form of clipping or frustrum culling is performed at this stage.


\textbf{Rasterization} uses a combination of Bresenham's algorithm \cite{koopman1987bresenham} and the Scanline algorithm \cite{lane1979generalized} to convert sets of three vertices into a wireframe triangle, and then into a filled triangle with interpolated colours. No form of anti-aliasing or subpixel interpolation is implemented. Fragments (candidate pixels) may overlap at this stage (i.e., have the same (x,y) position in the final image), but will be discarded if they are outside of the bounds of the canvas.

The \textbf{fragment shader} applies Blinn-Phong shading \cite{phong1975illumination,blinn1977models} to each fragment depending on the lighting configuration provided. The projected depths of each fragment are also used for depth testing, and only the closest fragments are kept (i.e., near parts of the object are rendered in favor of far away parts of the object if they occupy the same position in the render).

\textbf{Compositing} is the process whereby the pixel data is written into an image buffer (i.e. a 2D matrix with an RGB tuple at each element), and applied on top of a background. Shackled supports the generation of unicolour backgrounds, or two-color vertical gradient backgrounds. The image buffer is then encoded as a bitmap image. 

\subsection{Deployment}

Shackled's on-chain code consists of 13 smart contracts and libraries. The {deployment} of Shackled cost $0.258\Xi$ ($427$ USD at the time) and the deployment of Shackled Icons (a follow up tokenised art collection) cost $0.05\Xi$ ($80$ USD at the time). The cost of the follow up project was significantly less as the rendering engine code had already been deployed and could be used in a composable manner. The deployed and verified contracts can be found in Appendix~\ref{s:dataavail}. These projects tokenised the renders of Shackled as ERC-721A tokens (i.e., NFTs), demonstrating Shackled's commercial potential and viability in niche rendering use cases.

\begin{figure}[H]
    \centering
    \includegraphics[width=\textwidth]{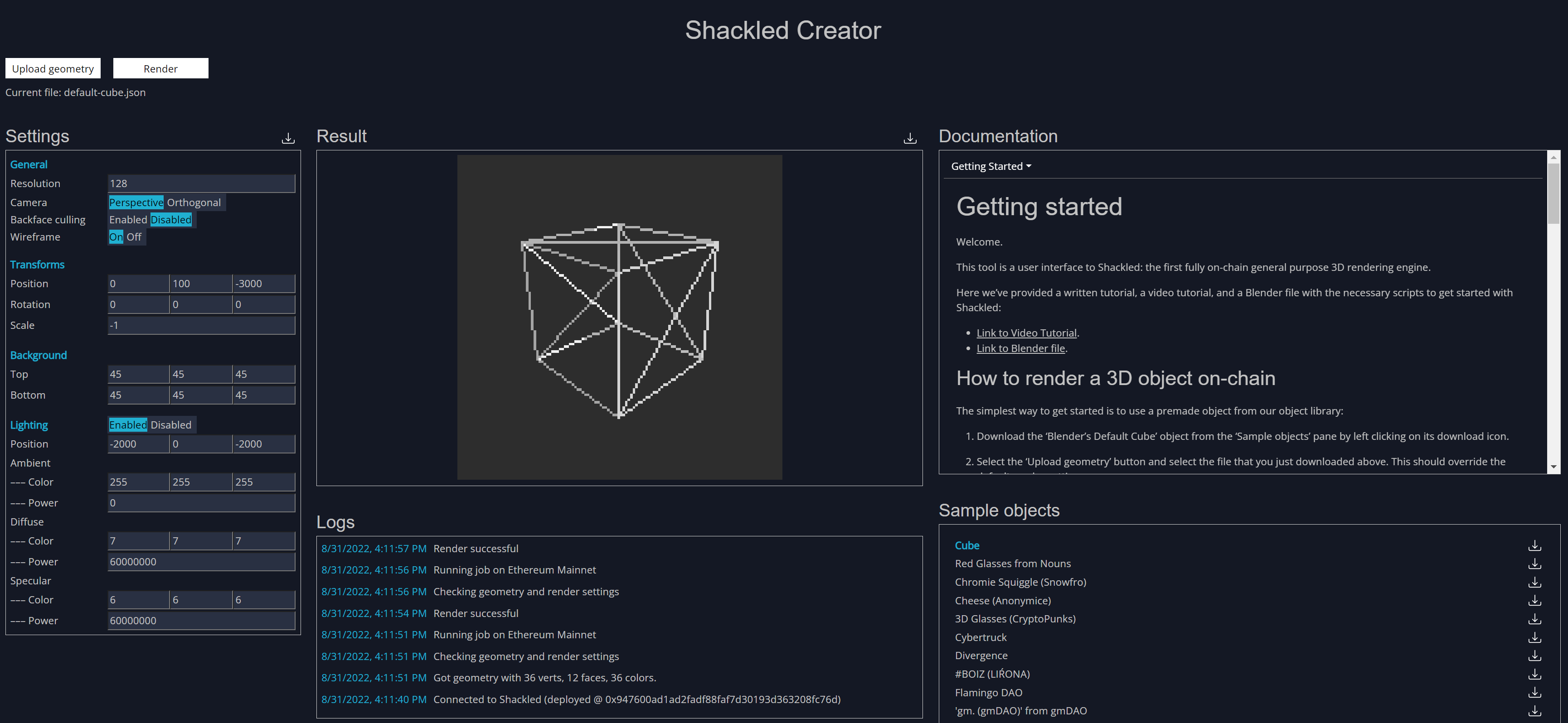}
    \caption{Shackled \textit{Creator}; a simple user interface to the Shackled rendering engine contracts. Note that rotation is provided as a transformation, however, the rotation matrix multiplications are calculated off-chain in the front-end. All other computations are performed on-chain.}
    \label{fig:creator}
\end{figure}

We also implemented, deployed, and provided a custom \textbf{user interface} (UI) to the Shackled smart contracts which enables a user to use Shackled for on-chain 3D rendering without any expertise in on-chain programming, Solidity, or smart contracts (see Figure~\ref{fig:creator}). This UI --- Shackled \textit{Creator} --- is free to use and publically available\footnote{Available at \href{https://shackled.spectra.art/\#/creator}{https://shackled.spectra.art/\#/creator.}}.


\section{Results and discussion}


\subsection{Computation scales quadratically with canvas size}

\begin{figure}[H]
    \centering
    \begin{subfigure}{0.715\textwidth}
        \setlength{\tabcolsep}{0pt}
        \begin{tabular}{ccccc}
            \includegraphics[width=0.2\linewidth]{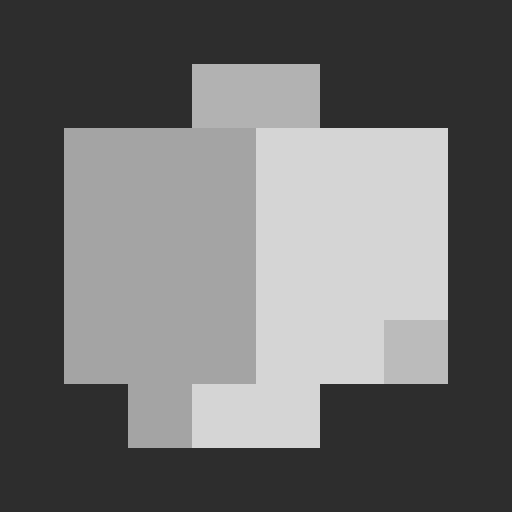}
            & 
            \includegraphics[width=0.2\linewidth]{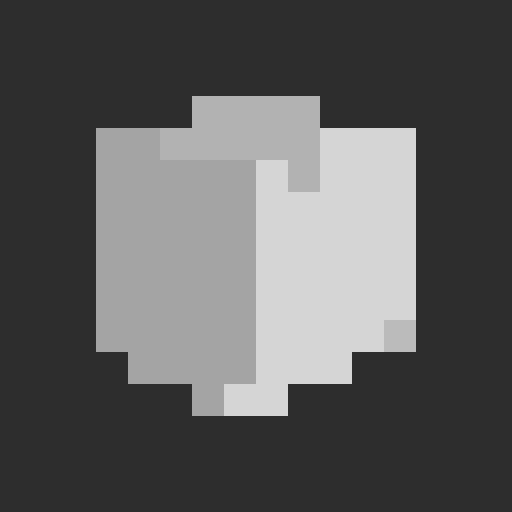}
            & 
            \includegraphics[width=0.2\linewidth]{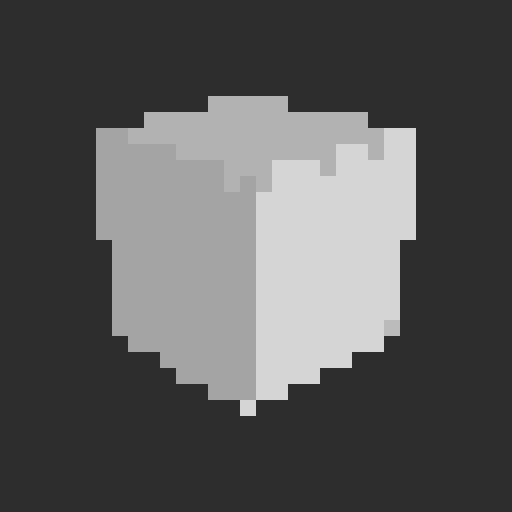}
            & 
            \includegraphics[width=0.2\linewidth]{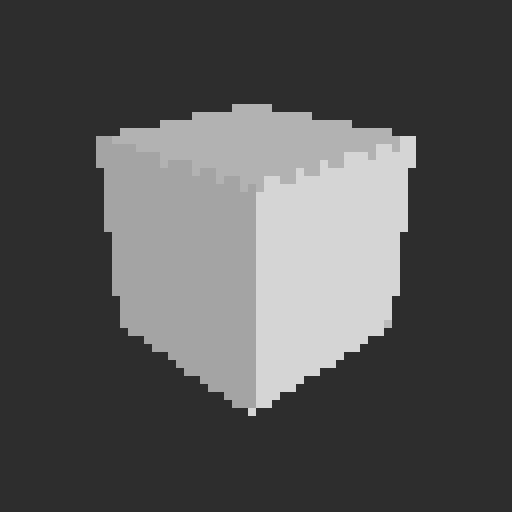}
            & 
            \includegraphics[width=0.2\linewidth]{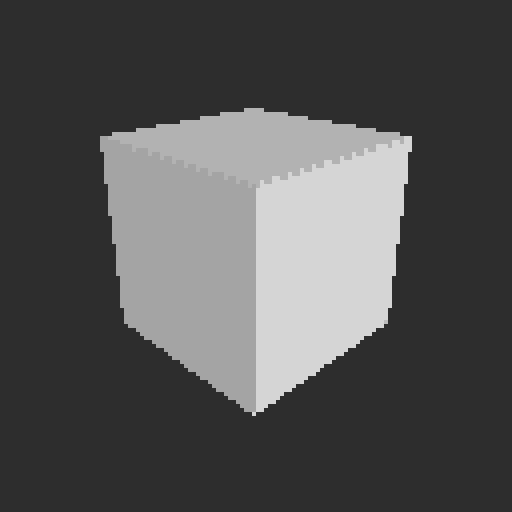}\\[-4pt]
            
            \includegraphics[width=0.2\linewidth]{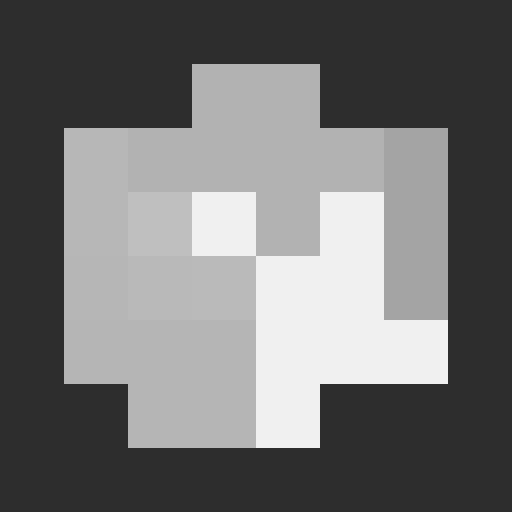}
            & 
            \includegraphics[width=0.2\linewidth]{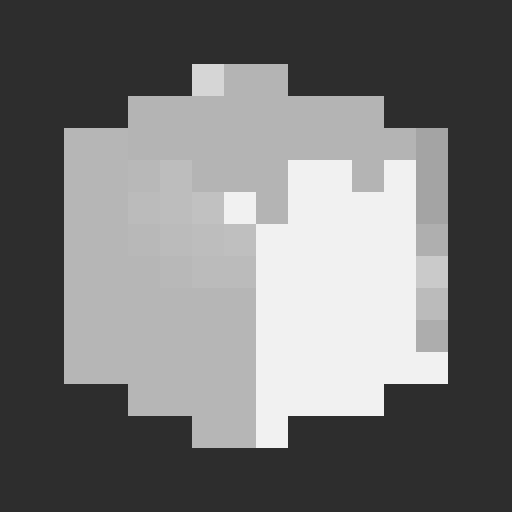}
            & 
            \includegraphics[width=0.2\linewidth]{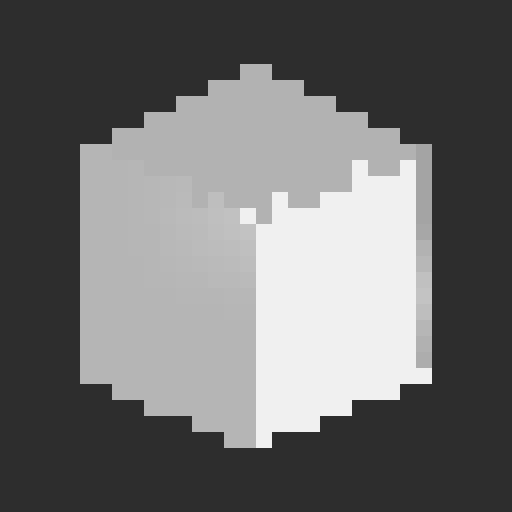}
            & 
            \includegraphics[width=0.2\linewidth]{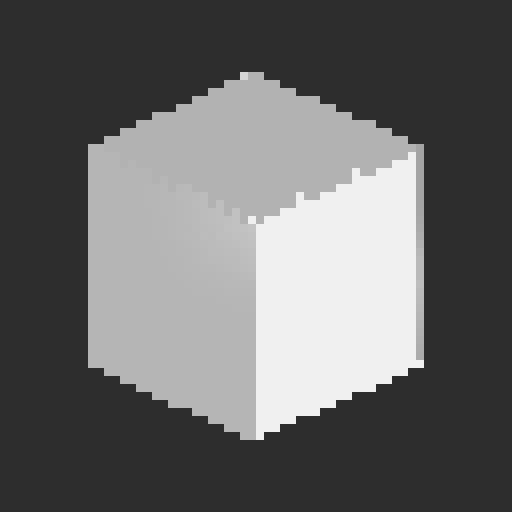}
            & 
            \includegraphics[width=0.2\linewidth]{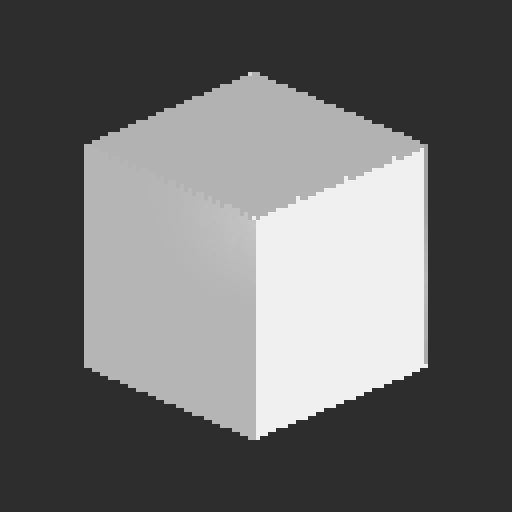}\\[-4pt]
            
        \end{tabular}%
    \end{subfigure}%
    \hfill
    \begin{subfigure}{0.284\textwidth}
        \includegraphics[width=1\linewidth]{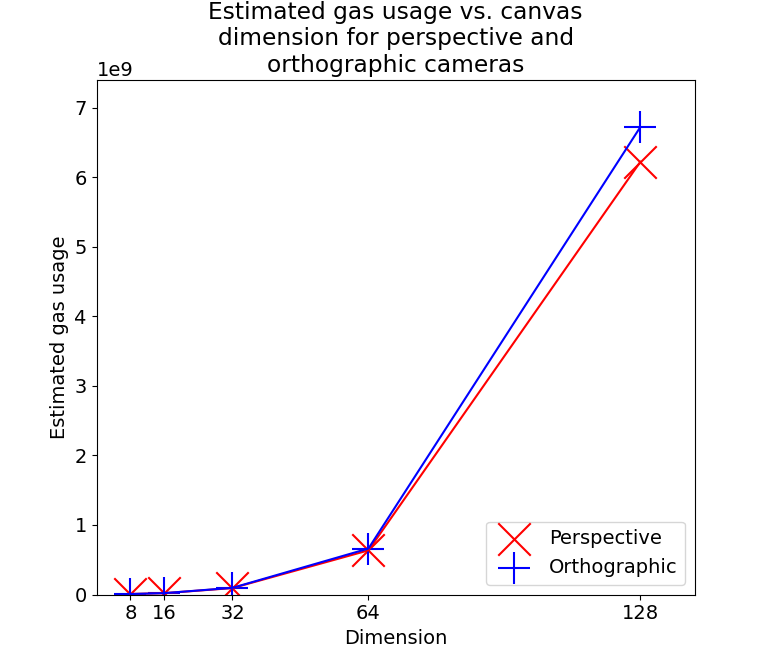}
    \end{subfigure}
    \caption{Rendering different cubes on-chain whilst varying canvas size and camera model parameters. \textbf{Left, top row}: varying the canvas size from 8 --- 128 pixels using a perspective camera. \textbf{Left, bottom row}: varying the canvas size from 8 --- 128 pixels using an orthographic camera. \textbf{Right}: plotting the gas estimates as a function of canvas size for each render using the perspective and orthographic camera settings.}
    \label{tab:canvas-size}
\end{figure}

We render the same object from a canvas size of $8\times8$ to $128\times128$ doubling the canvas size in each axis for each step. We use our gas estimation procedure to estimate the computational cost, and plot it in Figure~\ref{tab:canvas-size} (right). We compute two sets of results: one using the perspective camera projection model, and one using the orthographic camera projection model. The resultant renders are depicted in Figure~\ref{tab:canvas-size} (left).

We note that the computation cost scales quadratically with the canvas size, as expected due to the quadratic increase in fragments with canvas size. Additionally, the computation cost is invariant to the camera projection model. This quadratic scaling greatly limits the scalability of Shackled to larger renders, and potential improvements to alleviate this issue are suggested in Section~\ref{s:concfuture}.

\begin{figure}[H]    
    \centering
    \begin{subfigure}{0.45\textwidth}
        \setlength{\tabcolsep}{0pt}
        \begin{tabular}{cccc}
            \includegraphics[width=0.25\linewidth]{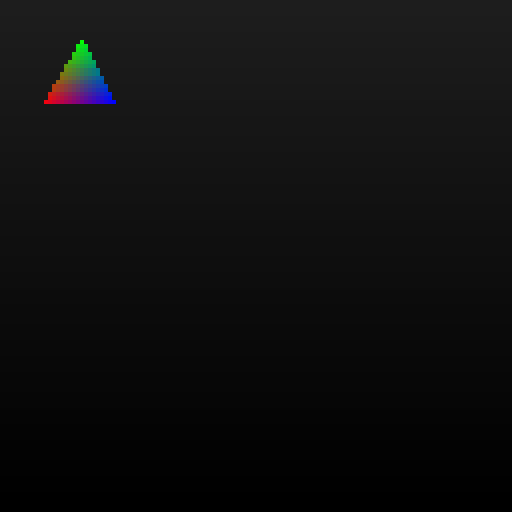}
            & 
            \includegraphics[width=0.25\linewidth]{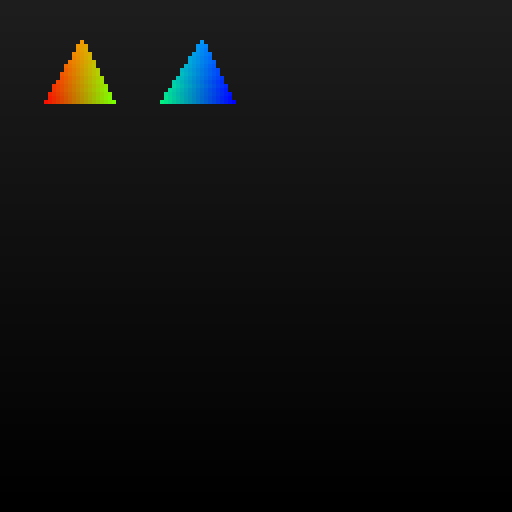}
            & 
            \includegraphics[width=0.25\linewidth]{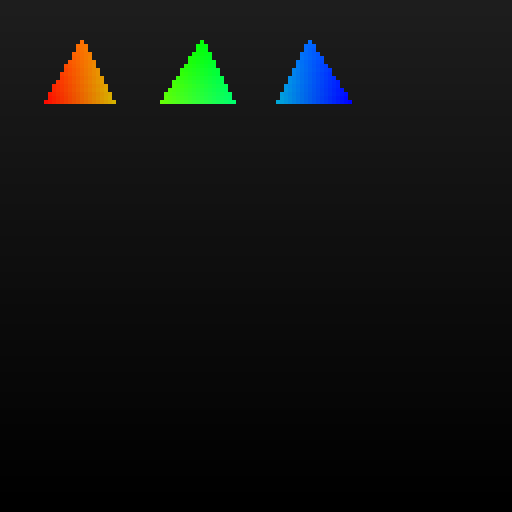}
            & 
            \includegraphics[width=0.25\linewidth]{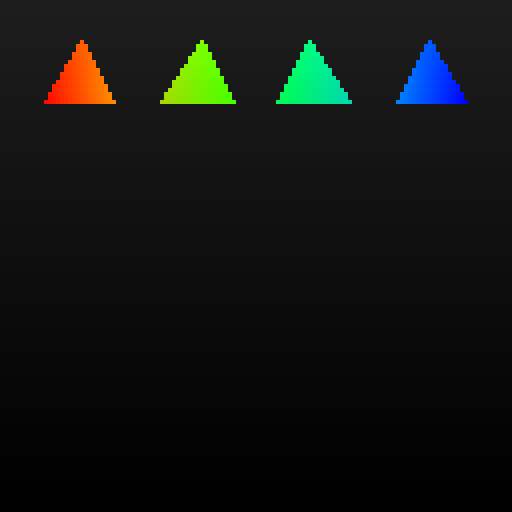}\\[-4pt]
            
            \includegraphics[width=0.25\linewidth]{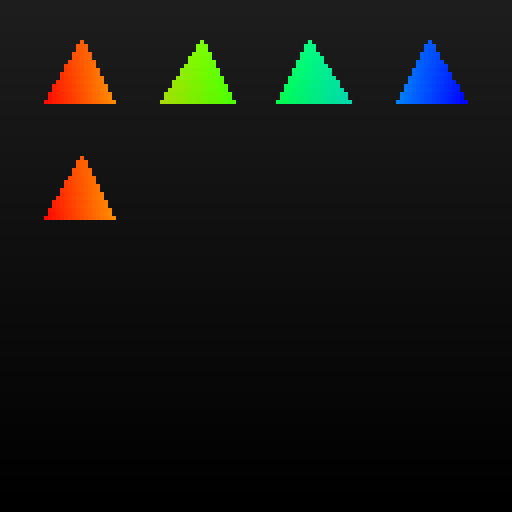}
            & 
            \includegraphics[width=0.25\linewidth]{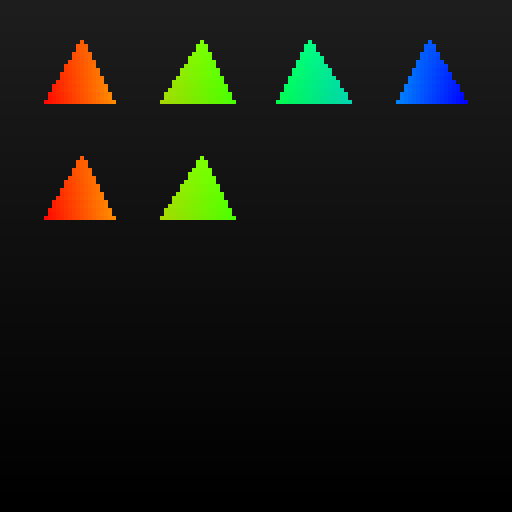}
            & 
            \includegraphics[width=0.25\linewidth]{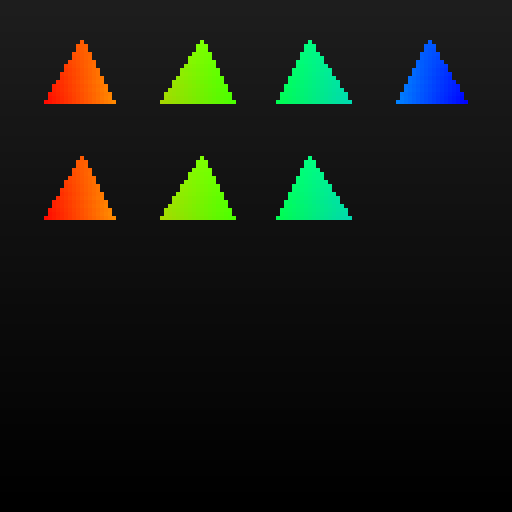}
            & 
            \includegraphics[width=0.25\linewidth]{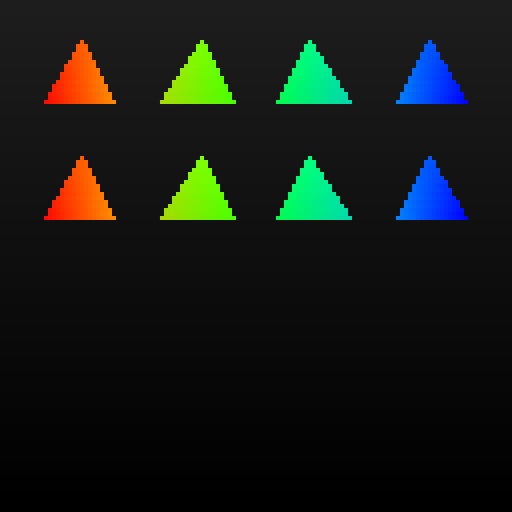}\\[-4pt]
            
            \includegraphics[width=0.25\linewidth]{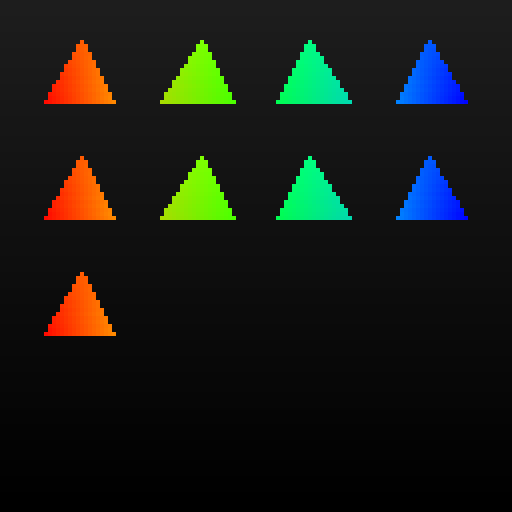}
            & 
            \includegraphics[width=0.25\linewidth]{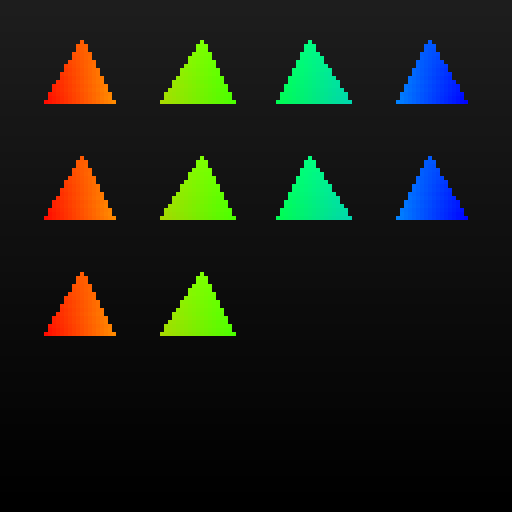}
            & 
            \includegraphics[width=0.25\linewidth]{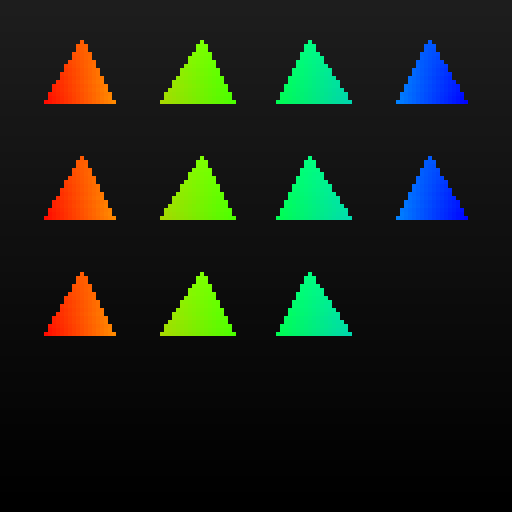}
            & 
            \includegraphics[width=0.25\linewidth]{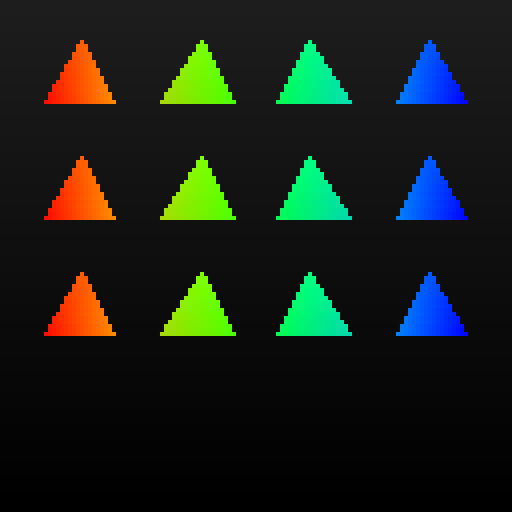}\\[-4pt]
            
            \includegraphics[width=0.25\linewidth]{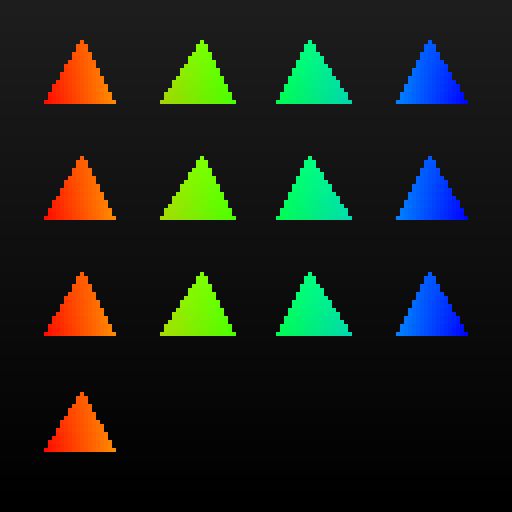}
            & 
            \includegraphics[width=0.25\linewidth]{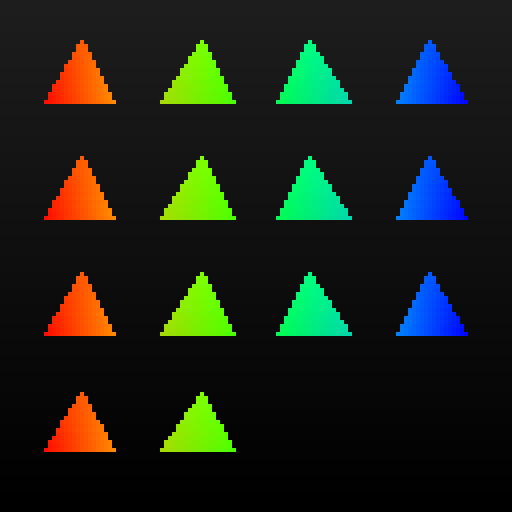}
            & 
            \includegraphics[width=0.25\linewidth]{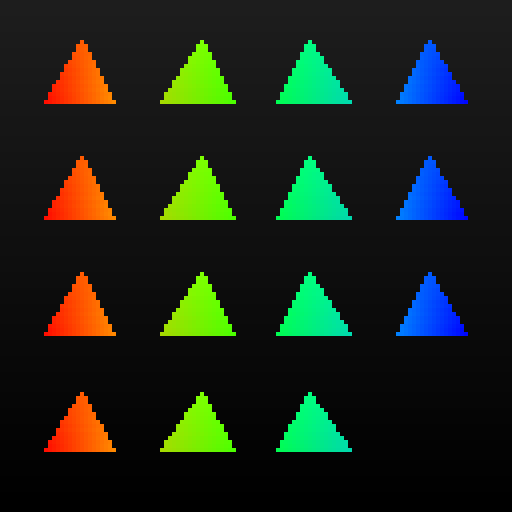}
            & 
            \includegraphics[width=0.25\linewidth]{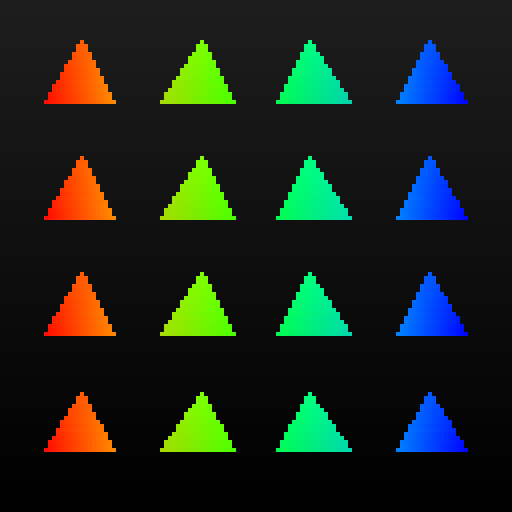}\\[-4pt]
        \end{tabular}%
    \end{subfigure}%
    \hfill
    \begin{subfigure}{0.55\textwidth}
        \includegraphics[width=1\linewidth]{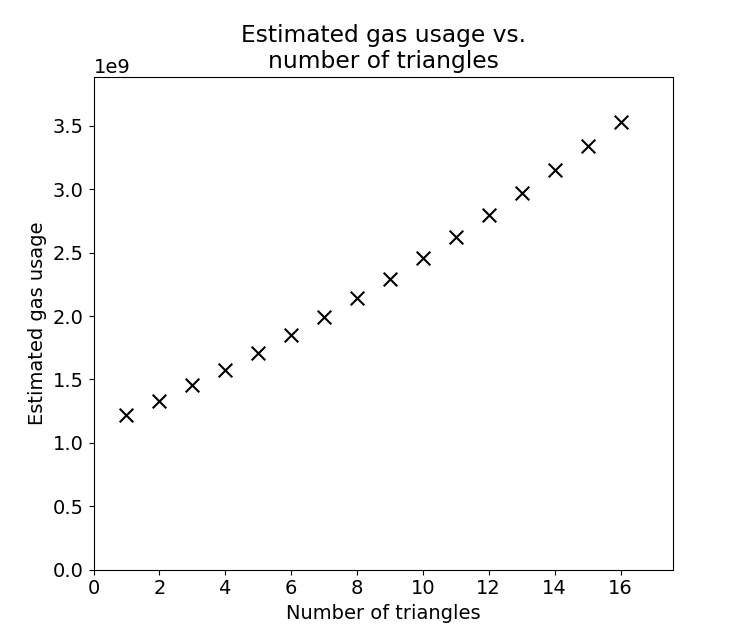}
    \end{subfigure}
    \caption{\textbf{Left}: the resultant renders from testing how the gas cost scales with the number of rendered pixels. The gas-pixel relationship is explored by rendering 1 --- 16 identical triangles, which result in a near-identical number of pixels per triangle in the final render. \textbf{Right}: the estimated gas usage increases approximately linearly with the number of triangles, and thus pixels. We note that the y-axis offset suggests that there is a constant amount of gas which is always `paid' regardless of the render.}
    \label{fig:increasingfrags}
\end{figure}

\begin{figure}[H]
    \centering
    \includegraphics[width=0.75\textwidth]{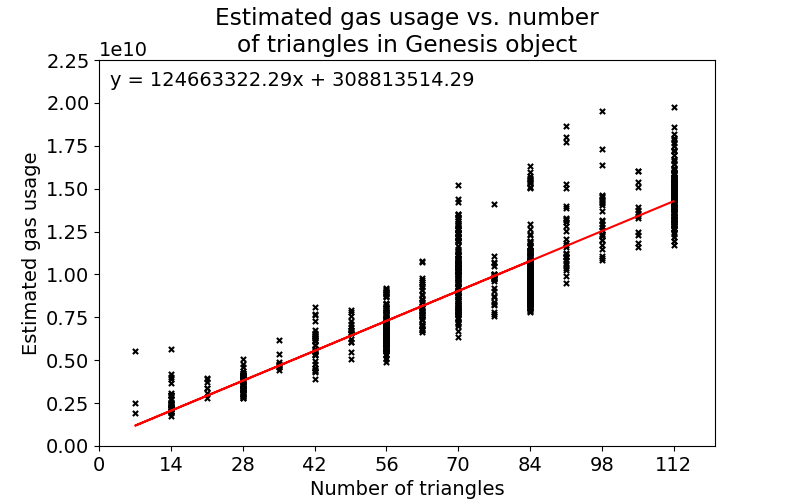}
    \caption{Estimated gas costs plotted against number of triangles in the rendered object, for every instance of the Shackled Genesis Dataset (see Table~\ref{fig:datasets}). A linear fit is applied using the NumPy package's {\fontfamily{qcr}\textit{polyfit}} function \cite{harris2020array}, and the result is plotted and displayed. Each triangular prism in each instance is comprised of $7$ triangles: $1$ for the front face, and $2$ triangles each for the $3$ sides of the prism (the prisms have no back), hence the $7$ unit separation in the x-axis.}
    \label{fig:increasingfragsgenesis}
\end{figure}

\subsection{Computation scales linearly with the number of triangles rendered}

We investigate the cost of rendering increasing numbers of pixels by incrementally rendering duplicates of an identical geometry (a single triangle), as illustrated in Figure~\ref{fig:increasingfrags}. We note that linearly increasing the number of triangles (and thus the number of fragments / pixels), results in a linear increase in the computational cost (as approximated by estimated gas). 

This allows Shackled to scale well to geometries with a large number of triangles, as long as the number of pixels on screen remains bounded. Indeed, this enables Shackled to successfully render an expressive range of objects on-chain, as illustrated in Table~\ref{fig:datasets}; the number of triangles in these objects range from $20$ (default cube) to $1198$ (Michaelangelo's David).

Additionally, we demonstrate a similar pattern using the entirety of the Shackled Genesis Dataset (samples illustrated in Table~\ref{fig:datasets}). The results are shown in Figure~\ref{fig:increasingfragsgenesis} and demonstrate again that there is an approximately linear relationship between the number of rendered pixels and the cost of computation (again approximated by estimated gas).



\subsection{Backface culling reduces computation significantly}

Furthermore, we investigate how backface culling \cite{blinn1993backface} --- an approach which analyses the normal vector of a triangle as it compares to the direction that the camera is pointing --- can greatly reduce the number of fragments that are computed by removing triangles that would never be visible in the final render. By reducing the number of fragments, we reduce the amount of required computation, and thus increase the capability of the rendering engine. 

We render the Shackled Common Graphics Objects Dataset with backface culling and without backface culling enabled. Without backface culling, the average gas estimate for rendering any object in the dataset is $1.54 \times 10^{11}$, compared to $4.39 \times 10^{10} $ with backface culling enabled; a decrease of $3.51$ times. We conclude that implementing on-chain backface culling greatly increases the capacity of Shackled to render complex objects, without any visible change to the output renders.

\section{Conclusion and future work}
\label{s:concfuture}

In this work, we have introduced and implemented Shackled, the first fully on-chain 3D rendering engine. We have benchmarked its efficiency using three custom datasets, demonstrated its use cases with respect to decentralised rendering, tokenised art, and as a blockchain-based platform for native 3D graphics processing. The datasets, source code, and a user interface for Shackled have been made available publically (see Appendix~\ref{s:dataavail}). Moreover, we have outlined the challenges associated with implementing complex algorithms (e.g. linear algebraic equations, trigonometry, and tensor operations) entirely in Solidity smart contracts.

In this early stage, Shackled illustrates only a fragment of the potential for fully on-chain algorithms --- particularly rendering engines. Candidates for \textbf{future work} that could improve Shackled include:
\begin{itemize}
    \item Removing rendering artifacts along straight lines and edges by implementing the digital differential analyzer algorithm in lieu of Bresenhams algorithm \cite{koopman1987bresenham}, as it is a better fit for interpolating with only integer arithmetic.
    \item Implementing on-chain rotation transformations using quaternions, Rodrigues' rotation, matrix products, or some other suitable 3D rotation algorithm.
    \item Taking proper advantage of the `embarassingly parallel' nature of 3D graphics rendering, e.g. rendering patches of a render simultaneously across nodes and compositing them together once complete.
\end{itemize}


\section*{Acknowledgements}




We would like to thank the \textit{entire team at Spectra} for working behind the scenes, sharing ideas which greatly improved the final Shackled renders, and for supporting the development of Shackled and its related projects. 

Moreover, running Shackled uses a relatively large gas-equivalent of computation (compared to traditional Ethereum smart contract functions). We would like to thank our partners at \textit{Alchemy} for specially raising the computation limits of an Ethereum read-call on their nodes so as to allow Shackled to operate. 

Finally, the creation of Shackled would not have been possible without the \textit{Spectra community}, and we would like to give a warm thanks to Doubtingtrev,
HollywoodMeta.eth,
Awesomerrificus, and
Max Bridgland for their work in directly managing, promoting, supporting, and moderating the community. 

In addition, we would like to thank Aubjectivity,
nodallydude,
Bon(g/j)e,
Paul Balaji,
Ntando Mhlungu,
Divirzion.eth,
a7111a.eth,
I2DT,
wakeupremember.eth,
Ian Orz,
Animechanic,
TriPoloski,
ZiK$\_$WaN,
NFTxDeFi,
Michael Slonim,
cearwylm,
Petter Rasmussen,
rpl.eth,
Dalst,
reechard.eth,
JP,
El Citadel,
Ralph Clayton,
Parker Thompson,
Ott Erlord, and
Jordysure for their support, feedback, and endorsements of Shackled. Fostering a positive community of folks that are excited about the technological advancement of on-chain art would not have been possible without you all.  
    

%
%
%
%


\bibliographystyle{alphaurl}
\bibliography{main}

\clearpage\section*{Appendix}

\renewcommand{\thesubsection}{\Alph{subsection}}

\subsection{Environmental considerations}

\label{s:carbon}

Shackled was developed during the Ethereum blockchain's proof-of-work era. The deployment and subsequent use of the two main contract libraries discussed in this work (Shackled Genesis and Shackled Icons) resulted in a total energy use equivalent to the emission of 54.6 metric tonnes of carbon dioxide, as measured by the Ethereum emissions calculator \textit{Carbon.FYI} \cite{carbonfyi2021}.

We are aware of the environmental impact of proof-of-work blockchains and have offset five times this amount of carbon dioxide --- 273 tonnes --- via the retirement of verified carbon credit units provided by \textit{Offsetra Limited}. Verra-accredited retired unit records are available at the following links:

\begin{enumerate}
    \item \href{https://registry.verra.org/myModule/rpt/myrpt.asp?r=206\&h=162735}{https://registry.verra.org/myModule/rpt/myrpt.asp?r=206\&h=162735}.
    \item \href{https://registry.verra.org/myModule/rpt/myrpt.asp?r=206\&h=166285}{https://registry.verra.org/myModule/rpt/myrpt.asp?r=206\&h=166285}.
\end{enumerate}

\subsection{Data availability and reproducibility}
\label{s:dataavail}


Data, tools, and code are available at the following links:

\begin{itemize}
    \item \href{https://shackled.spectra.art/}{Shackled website}.
    \item \href{https://shackled.spectra.art/\#/creator}{Shackled no-code user interface website}.
    \item \href{https://github.com/ike-ike-ike/shackled-paper}{GitHub repository for accessing datasets.}
    \item \href{https://etherscan.io/address/0x947600ad1ad2fadf88faf7d30193d363208fc76d}{Verified smart contracts for Shackled on Etherscan.}
\end{itemize}




\subsection{The first on-chain render computed by Shackled}

\begin{figure}
    \centering
    \includegraphics[width=\textwidth]{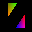}
    \caption{The first on-chain render computed by Shackled. We believe this to be \textbf{the first on-chain 3D render ever computed on Ethereum}. The triangles naturally appear 2D in the render, but are defined in 3D space by the points (-1,0,0), (-1,2,0), (0,2,0) and (1,0,5), (1,-2,5), (0,-2,5) respectively.}
    \label{fig:firstrender}
\end{figure}

As Shackled is the first on-chain 3D rendering engine, we believe its first output to be the first on-chain 3D render. We have embedded that render into this paper for posterity (Figure~\ref{fig:firstrender}).

%

\end{document}